\documentclass[twocolumn]{aastex631}
\pdfoutput=1 
\usepackage{amsmath,amstext,amssymb,mathtools,graphicx,color}

\usepackage{graphicx}	
\usepackage{amsmath}	
\usepackage{amssymb}	
\usepackage{xcolor}

\received{June 1, 2019}
\revised{January 10, 2019}
\accepted{\today}
\submitjournal{ApJ}

\shorttitle{Remnant Masses}
\shortauthors{Fryer et al.}

\graphicspath{{./}{figures/}}

\begin{document}

\title{The Effect of Supernova Convection On Neutron Star and Black Hole Masses}

\correspondingauthor{Chris L. Fryer}
\email{fryer@lanl.gov}

\author[0000-0003-2624-0056]{Chris L. Fryer}
\affiliation{Center for Theoretical Astrophysics, Los Alamos National Laboratory, Los Alamos, NM, 87545, USA}
\affiliation{Computer, Computational, and Statistical Sciences Division, Los Alamos National Laboratory, Los Alamos, NM, 87545, USA}
\affiliation{The University of Arizona, Tucson, AZ 85721, USA}
\affiliation{Department of Physics and Astronomy, The University of New Mexico, Albuquerque, NM 87131, USA}
\affiliation{The George Washington University, Washington, DC 20052, USA}
\author[0000-0002-6105-6492]{Aleksandra Olejak}
\affiliation{Nicolaus Copernicus Astronomical Center, Polish Academy of Sciences, 18 Bartycka Street, 00-716 Warsaw, Poland}
\author{Krzysztof Belczynski}
\affiliation{Nicolaus Copernicus Astronomical Center, Polish Academy of Sciences, 18 Bartycka Street, 00-716 Warsaw, Poland}



\begin{abstract}

Our understanding of the convective-engine paradigm driving core-collapse supernovae has been used for 2 decades to predict the remnant mass distribution from stellar collapse.  These predictions improve as our understanding of this engine increases.  In this paper, we review our current understanding of convection (in particular, the growth rate of convection) in stellar collapse and study its effect on the remnant mass distribution.  We show how the depth of the mass gap between neutron stars and black holes can help probe this convective growth.  We include a study of the effects of stochasticity in both the stellar structure and the convective seeds caused by stellar burning.  We study the role of rotation and its effect on the pair-instability mass gap.  Under the paradigm limiting stellar rotation to those stars in tight binaries, we determine the effect of rotation on the remnant mass distribution.

\end{abstract}



\section{Introduction} \label{sec:intro}

Early calculations of compact remnant masses were based solely on our understanding of stellar structure, guided by nucleosynthetic yields and the use of a piston explosion model~\citep{1996ApJ...457..834T}.  Many of the recent remnant-mass distribution calculations have focused on improving 1-dimensional explosion calculations to capture the engine more accurately than the piston model.  However, the convective-engine behind supernova explosions is an intrinsically multi-dimensional process and incorporating this physics is difficult in a 1-dimensional simulation.  The explosion is typically driven by artificially altering the energy deposition in a 1-dimensional code.  A large fraction of these models altered either the neutrino luminosity or absorption to increase the energy deposited~\citep{2006ApJ...637..415F,2010A&A...517A..80F,2012ApJ...757...69U,2015ApJ...806..275P,2016AJ....152...41P,2016ApJ...818..124E}.  These models mimic the enhanced neutrino luminosity and heating from convection within the proto-neutron star~\citep{1996ApJ...473L.111K}.  Other models have implemented 1-dimensional mixing above the proto-neutron star (which is believed to be much more important than the proto-stellar mixing)  to drive explosions~\citep{2018ApJ...856...63F,2020ApJ...890..127C}.   All of these models are simply recipes to mimic in 1-dimension the intrinsically multi-dimensional supernova engine.

To understand the difficulties in producing an accurate explosion in 1-dimension, we must better understand the current paradigm for the engine behind core-collapse supernovae.   \cite{herant94} found that convection above the proto-neutron star can both enhance the conversion of internal energy produced in the bounce and reduce the pressure of the infalling material.  Since these results, an increasing number of simulations have produced explosions under this paradigm, and it has slowly gained traction to become the leading theory model for the supernova engine~\citep{2007ApJ...659.1438F,2014ApJ...786...83T,2015ApJ...807L..31L,2015ApJ...808L..42M,2018SSRv..214...33B,2021arXiv210704617F}.  Given that convection is key to the supernova engine, it would appear that the models including mixing would be more predictive than those that simply focus on the neutrino physics.  But, as we shall discover in this paper, different 1-dimensional prescriptions for mixing produce different results.  In addition, these prescriptions do not capture the large-scale flows seen multi-dimensional models that will affect the final remnant mass distribution.  

Unfortunately, multi-dimensional models also do not provide a final answer.  The growth of convective instabilities in multi-dimensional models tends to be slower than the growth times predicted by turbulence models~\citep{2007ApJ...659.1438F}.  The complexity (multi-physics) and scale of the core-collapse supernova engine prevents scientists from conducting fully-resolved simulations of this convection.  The resulting numerical viscosity in most calculations damp out the growth of convection, delaying the development of strong convection.  Because of the transient nature of supernovae, the timing of this convective growth can have a large quantitative impact on the strength of the supernova engine~\citep{2007ApJ...659.1438F}.

On top of the numerical constraints in the engine itself, convective instabilities in the stellar progenitor also play a role in the growth of convection in the supernova engine.  Early calculations justified perturbations in their numerical setup by arguing that these perturbations were on par with the convective instabilities seen in stellar convection~\citep{herant94,2004ApJ...601..391F,2007ApJ...659.1438F}.  More extreme initial asymmeties caused by vigorous silicon convection were proposed as the source for strong neuton star kicks~\citep{1996PhRvL..76..352B,2004ApJ...601L.175F}.  But it was not until modelers coupled multi-dimensional simulations of the pre-collapse star with multi-dimensional engine models that the community has begun to accept the importance of stellar convection in accelerating the engine~\citep{2013ApJ...778L...7C,2021arXiv210704617F}.  Unfortunateley, these pre-collapse models are also under-resolved and poorly understood, further preventing a quantitative, first-principles model for the supernova engine.

Calculating a remnant mass distribution of more massive stars depends on a different set of collapse properties, in particular, the angular momentum.  If the collapsing star has considerable angular momentum and the proto-neutron star develops strong magnetic fields, a magnetar engine can drive an explosion.  These explosions do not depend upon convection and will produce a different remnant mass than normal supernovae.  Further, if a rapidly rotation star collapses to form a black hole, the resultant black hole accretion disk can drive a jet that is both believed to produce long-duration gamma-ray bursts~\citep{1993ApJ...405..273W,1999ApJ...524..262M,2001ApJ...550..410M} and eject much of the stellar envelope.  Like our standard supernova models, simulations of magnetic field and jet production are under-resolved and are not yet predictive.  Also paralleling the convective engine paradigm, the angular momentum of the collapsing core depends upon models of stellar convection and their uncertainties.  As such, it is currently beyond our computational power to model this more exotic engine and its effect on the remnant population.  However, by making assumptions on the nature of this rotation, we can estimate the role rotating engines play on the compact remnant distribution.

All of these obstacles in developing a quantitative core-collapse engine model has led some groups to instead follow an approach that leverages knowledge gained from multi-dimensional models, 1-dimensional solutions and an understanding of turbulence growth to develop an analytic model for the remnant mass distribution~\citep{2001ApJ...554..548F,2012ApJ...749...91F,2021MNRAS.500.5393G}.  In this paper, we apply a range of analytic studies of turbulence growth to core-collapse conditions to develop our intuition for the convective engine (Section~\ref{sec:mixing}).  With this understanding of the physics behind this convective engine and its resultant uncertainties, we develop a range of mass distributions for neutron stars and low-mass black holes from supernovae (Section~\ref{sec:remnant}).  The effect of rotation on the remnant mass is discussed in Section~\ref{sec:rotation}.  With these models implemented into binary population synthesis models, we can then predict mass distributions in observed binary systems, including the merging binaries observed through gravitational waves (Section~\ref{sec:binarybh}).  We conclude with a summary of the primary findings of this paper.

\section{Understanding Prescriptions of Convection}
\label{sec:mixing}

Driven both by theoretical studies and observational insight, the basic paradigm behind the core collapse engine has evolved with time.  For type Ib/c and II supernovae, the energy source for the explosion is derived from the potential energy released when the core of a massive star collapses to nuclear densities.  Although roughly $10^{53} {\rm erg}$ is released in the collapse, tapping just 1\% of this energy has proven difficult.  A clue into the nature of this explosion lies in the extensive mixing observed in the innermost ejecta (iron peak elements) of SN 1987A.  Convection above the newly formed proto-neutron star both provides a means to extract the potential energy released and a way to prevent a pile-up of infalling material that would make it increasingly difficult to drive an explosion~\citep[for reviews, see][]{herant94,2007ApJ...659.1438F,2021ARep...65..937F}.

In this paper, we focus on the uncertainties in this convection and its growth, ultimately studying its effect on compact remnant masses.  Herant and collaborators initially proposed Rayleigh-Taylor instabilities as the dominant driver behind convection~\citep{herant94} and multiple studies have confirmed that the growth time of this convection is very rapid and it can dominant the matter motion above the proto-neutron star~\citep{2004ApJ...601..391F,2007ApJ...659.1438F,2013ApJ...778L...7C,2018MNRAS.477.3091V,2019MNRAS.482..351V,2021arXiv210704617F}.  The entropy structure set by the stalled shock and infalling stellar material has some of the same properties of accreting neutron stars~\citep{1991ApJ...376..234H,1996ApJ...460..801F}.  The entropy in the shock decreases with decreasing shock velocity and lower accretion rates.  As the bounce shock propagates outward, the shock velocity (and, hence, the entropy at the shock) decreases.  In both supernovae and neutron star accretion, the entropy profile set by the outward propagation of the shock is susceptible to Rayleigh Taylor instabilities.  Growth timescales of the Rayleigh Taylor instability is extremely rapid~\citep{2007ApJ...659.1438F} and, as we shall show here, strong turbulent convection can develop in the first 3-5\,ms after bounce.  Although other forms of turbulence exist, e.g. standing accretion shock instabilities~\citep{2003ApJ...584..971B,2021MNRAS.503.3617A}, the fast growth of the Rayleigh-Taylor instability suggests it will dominate the turbulent convection in many supernova scenarios.  We will focus on Rayleigh Taylor convection in this paper.

Rayleigh-Taylor instabilities occur when there is an inversion in material properties under an acceleration (such as gravity), i.e. high-entropy material lies beneath low-entropy material (e.g. low-density fluid beneath higher-density material).  Such a density/entropy gradient occurs in many simulations (the extent depends upon the equation of state) of the collapse bounce of a massive core.  But for convection to occur, initial perturbations in the entropy or "seeds" to this convection must exist (presumably from asymmetries in the collapsing core).  In addition, the growth time depends on the viscosity of the material~\citep{2007PhRvE..76d6313D,2011PhPl...18b2109B,2012PhFl...24g4107R}.  The numerical viscosity in the under-resolved core-collapse calculations is many orders of magnitude higher than Spitzer viscosity from electron/ion transport~\citep{1953PhRv...89..977S}, neutrino viscosity~\citep{2007ApJ...659.1438F}, or effective viscosities from magnetic fields~\citep{1976PASJ...28..451N,2015MmSAI..86...89M,2019PhRvF...4h3701P}.

Unfortunately, the complexity of the supernova convection problem prevents state-of-the-art calculations from fully resolving the viscous scale.    Most of the supernova calculations using large eddy simulations which do not capture the small scale turbulence.  The numerical viscosity exceeds any physical viscosity, delaying the growth of turbulence.  It has been argued that most of the energy critical to driving an explosion will ultimately reside in the large-scale flows in an attempt to justify these under-resolved calculations.  While true, this resolution is often critical in determining the growth time of these large-scale flows.  The dynamic nature of core-collapse supernovae (short evolutionary timescales) requires accurate calculations of the growth time of this convection to  develop a quantitative picture of these explosions.  Most calculations in the literature do not include any representation of the sub-grid turbulence.  Sub-grid models could be used to understand the turbulent evolution below the resolution scale of the simulations.  Reynolds-averaged Navier-Stokes sub-grid models are commonplace in many engineering applications but such models also have their limitations and are typically tuned to match data for specific applications.  In this section, we will study these sub-grid models in more detail.

Although further studies of the supernova engine in 3-dimensions are critical to increasing our understanding of supernovae, if you combine these uncertainties in mixing with the vast array of additional physical uncertainties, it becomes clear that a quantitative, first-principles model of the core-collapse engine remains an ultimate goal and not a near-term goal to understanding the compact remnant population.  Understanding this convective supernova engine requires a two-prong approach:  both an ever-improving set of 2- and 3-dimensional core-collapse models and a set of 1-dimensional models that are guided by these studies to extract new insight from neutron star remnant populations.  

1-dimensional models of supernovae have been used for many decades, allowing scientists to study the fate of collapsing cores for a broad range of supernova progenitors~\citep{1986NYASA.470..267W,1988ApJ...334..909M,1988PhR...163...63W,1999ApJ...516..892F,2001ApJ...554..548F,2006ApJ...637..415F,2010A&A...517A..80F,2012ApJ...757...69U,2015ApJ...806..275P,2016AJ....152...41P,2016ApJ...818..124E,2018ApJ...856...63F,2020ApJ...890..127C}.  Initially, explosions were initiated either through piston or thermal bombs, driving strong explosions that matched observed supernova energies.  With these studies, scientists began studying additional supernova observables like nucleosynthetic yields, using these yields to probe characteristics of the progenitor.  Newer studies have instead focused on varying the efficiency of neutrino energy deposition to drive an explosion~\citep{2006ApJ...637..415F,2010A&A...517A..80F,2012ApJ...757...69U,2015ApJ...806..275P,2016AJ....152...41P,2016ApJ...818..124E}.  These later studies could potentially be used to probe aspects of the neutrino physics.  Finally, another set of calculations have focused on the convective engine and parameterized the convection to drive the explosion~\citep{2001ApJ...554..548F,2018ApJ...856...63F}.  The most recent of these have used physics-based sub-grid mixing models for their explosions~\citep{2019ApJ...887...43M,2020ApJ...890..127C}. 

These studies focused on only one or two instantiations of these recipes to produce a suite of models (often calibrating to SN 1987A and an assumption for this supernova's progenitor).  Such studies are ideally suited to produce results for population studies, but we must bear in mind that, without a complete understanding of the physics, there are broad uncertainties in the theoretical predictions.  Focusing on just the convection, we can begin to understand (and hopefully characterize) this range simply by looking at the different predictions from different sub-grid mixing algorithms, applying them to the profiles of post-bounce.  For this study, we use a mixing-length theory (MLT) approach commonly used in astrophysics and a Reynolds-averaged Navier Stokes (RANS) approach based on models developed in the engineering community~\cite{2009JTurb..10...13L}.  In a sense, MLT is a very particular RANS model, but it differs enough that it should be considered separately.

The MLT solution we apply is identical to the one described in detail by~\cite{2020ApJ...890..127C}.  If we focus on the source terms (neglect the flux terms), the turbulent velocity  growth is described by:
\begin{equation}
    \frac{\partial (\rho v_{\rm turb}^2)}{\partial t} + \, {\rm Flux \; Terms} = \rho v_{\rm turb} \omega^2_{\rm BV} \Lambda_{\rm mix} - \rho v_{\rm turb}^3/\Lambda_{\rm mix}.
\end{equation}
where $\rho$ is the density and $v_{\rm turb}$ is the turbulent velocity.  In this calculation, $\omega_{\rm BV}$ is the Brunt-V\"ais\"al\"a frequency:
\begin{equation}
    \omega_{\rm BV}^2 = - g_{\rm eff} (1/\rho \partial \rho/\partial r - (\rho c_s^2)^{-1} \partial P/\partial r )
\end{equation}
where $c_s$ is the sound speed, $P$ is the pressure and $g_{\rm eff}=-\partial \Phi/\partial r + v_r \partial v_r/\partial r$ is the effective gravity with $\Phi$ is the gravitational potential and $v_r$ is the radial velocity.  Finally, $\Lambda_{\rm mix}$ is the mixing length set to a fraction $\alpha_\Lambda$ of the pressure scale height.

Typical RANS solutions are very similar, but the notation, and driving terms are different.  For this solution, we use the turbulent kinetic energy ($\widetilde{k}$).  A simple driving term is in the RANS formulism (taking only the leading-order terms) is~\cite{2009JTurb..10...13L}:
\begin{equation}
    \frac{\partial \rho \widetilde{k}}{\partial t} = -(\rho \widetilde{U}_3 \widetilde{k})_{,3} + a_3 \frac{\partial P}{\partial r} - R_{33} \widetilde{U}_{3,3} - 1/2 R_{ii3,3} 
\end{equation}
where $a_3$ is:
\begin{equation}
    a_3 = h C_t \widetilde{k}^{1/2}
\end{equation}
where, following \cite{2009JTurb..10...13L}, we set $C_t = 0.39$ (constant defining the turbulent viscosity) and $h$ is the pressure scale height.  In these equations, $\rho$ is the density, $\widetilde{U}_3$ is the average radial velocity from our 1-dimensional calculations, $R_{ij}$ is the Favre Reynolds stress described in equation 10 of \cite{2009JTurb..10...13L}, $\widetilde{k}$ is the turbulent kinetic energy described in equation 11 of \cite{2009JTurb..10...13L}.  The $3$ in these equations refers to the radial direction and, because we only have radial terms in our 1-dimensional calculation, these are the only terms considered.

\begin{figure}
\includegraphics[scale=0.5,angle=0]{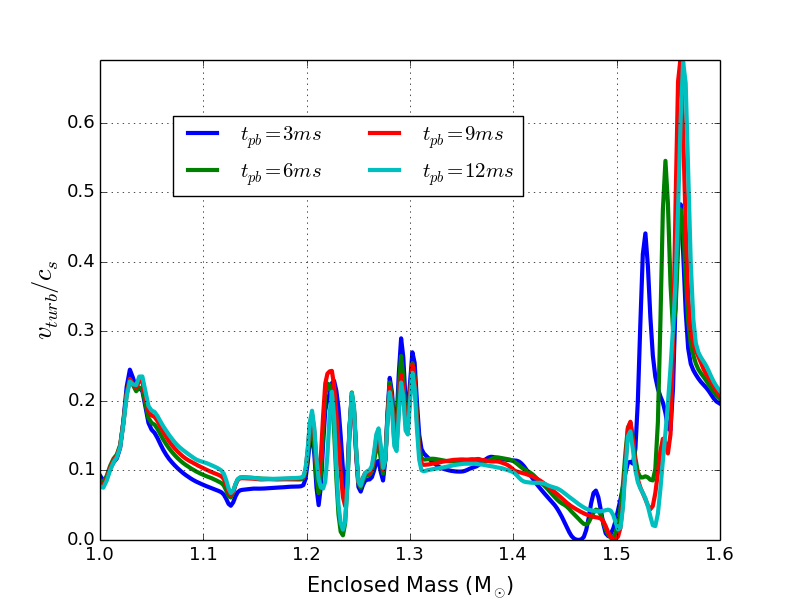}
\includegraphics[scale=0.4,angle=0]{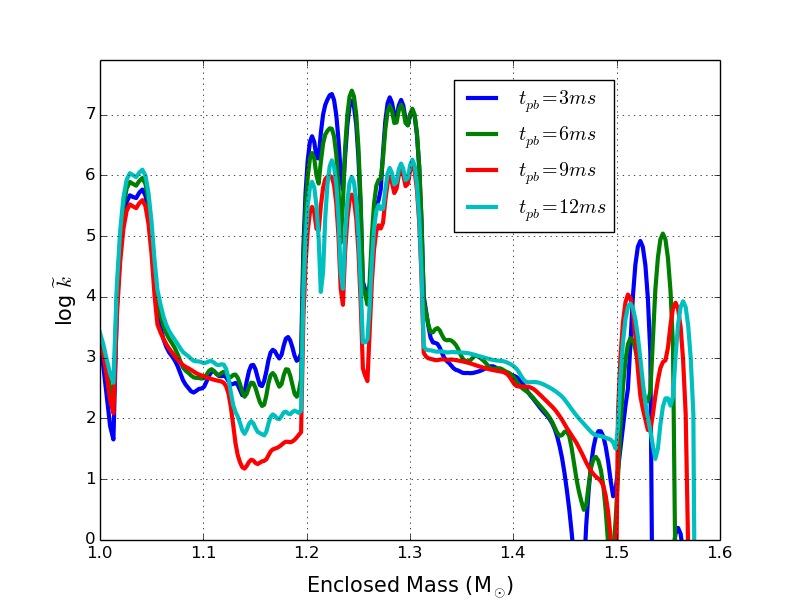}
\caption{Growth of turbulence, sound speed from mixing-length theory~\citep{1958ZA.....46..257K}) and turbulent energy from a Reynolds Averaged Navier Stokes solution~\cite{2009JTurb..10...13L} for a 15\,M$_\odot$ progenitor 3, 6, 9, and 12\,ms after bounce.  Strong turbulence develops in less than 3\,ms.} 
\label{fig:mixcomp}
\end{figure}

To compare the growth of these different instabilities, we take massive star progenitors and collapse and bounce the stars to determine post-bounce conditions.  For our study, we use both 15 and 25\,M$_\odot$ progenitor stars modeled with the Kepler code~\citep{2017ascl.soft02007W}.  These stars are then collapsed using our 1-dimensional core-collapse code~\citep{1999ApJ...516..892F}.  This code includes a simple neutrino transport (gray flux-limited diffusion) for each of the 3 neutrinos species, a dense nuclear equation of state, and general relativity assuming spherical symmetry.  With this code, we can model the collapse and bounce of the massive star.   

\begin{figure}
\includegraphics[scale=0.45,angle=0]{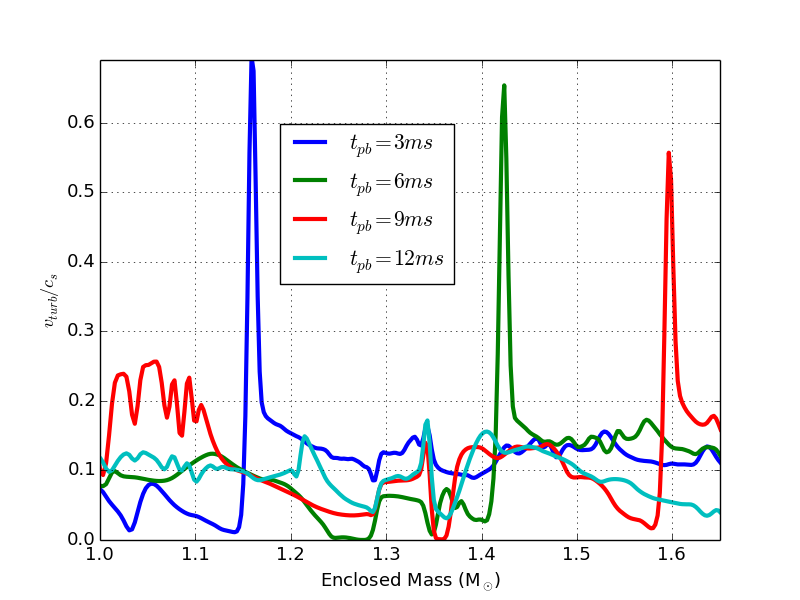}
\includegraphics[scale=0.45,angle=0]{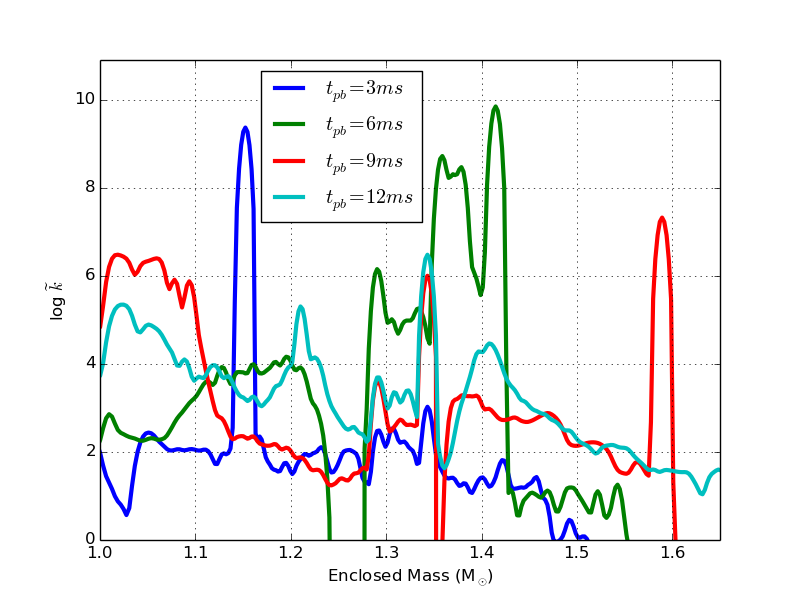}
\caption{Growth of turbulence, sound speed from mixing-length theory~\citep{1958ZA.....46..257K}) and turbulent energy from a Reynolds Averaged Navier Stokes solution~\cite{2009JTurb..10...13L} for a 25\,M$_\odot$ progenitor 3, 6, 9, and 12\,ms after bounce.  Strong turbulence develops in less than 3\,ms.} 
\label{fig:mixcomp2}
\end{figure}

We use the post-bounce structure from these simulations to test the growth times of the instabilities.  Figures~\ref{fig:mixcomp} and \ref{fig:mixcomp2} show the growth of turbulence at 4 different times post-bounce for 15 and 25\,M$_\odot$ progenitors respectively.  The top panel shows the turbulent velocity using our MLT approximation using a relatively low value for the mixing length ($\alpha_\Lambda=0.1$).  The bottom panels show the turbulent energy in our RANS formalism.  Here we have taken the structure of the collapsed star at a 3, 6, 9 and 12\,ms period post-bounce and, keeping these conditions fixed, determined the growth of instabilities over a 3\,ms period.  Both of our subgrid models agree that Rayleigh-Taylor instabilities are strong and grow quickly in these post-bounce conditions (something not seen in the under-resolved multi-dimensional calculations).  The models also agree on the regions where the instability grows.  But they do not agree on which regions dominate this growth.  Although analytical derivations show that our multi-dimensional models are not capturing this physics entirely, we still can not predict exactly the growth of turbulence.  For this paper, we allow the growth time to vary to determine its effect on the compact remnant mass distribution.

At this time, we can not predict the growth of convection exactly with either resolved multi-dimensional models or reduced-term analytic models.  But we can understand the trends in this convective growth and use it to guide our remnant-mass models.  In the next section, we show how measurements of the remnant mass distribution can help validate our convection models.

\section{Convection and the Remnant Mass Distribution}
\label{sec:remnant}

To determine how the growth of convection alters the remnant mass, we have altered the prescription used in the entropy-driven calculations of remnant mass distributions~\citep{2001ApJ...554..548F, 2012ApJ...749...91F}.  These calculations assume a convective region whose entropy evolves with time (mimicking the convection).  In \cite{2012ApJ...749...91F}, the fast- and slow-convection models were developed to demonstrate the extremes allowed under the convective model.  But the true answer likely lies in between these values.  In this section, we produce a simple prescription to vary the remnant mass due to differences in the convective growth time.  

This prescription does not include the possibility that massive star evolution may produce some stochasticity into the supernova explosion mechanism.  Many stellar evolution calculations have produced wild variations in the core structure at collapse that then produce large variations in the final remnant mass~\cite{2021A&A...656A..58L,2021A&A...645A...5S}.  Whether or not these wild variations in stellar-structure are real (vs. numerical artifacts), the fact that the nature of the asymmetries in the collapsing core can alter the growth of convection suggests that at least some stochasticity exists in the models.  We include remnant-mass prescriptions including this stochasticity at varying levels to determine whether individual stellar systems or population studies can be used to place limits on the level of this stochasticity.

\subsection{Growth Time Variations}

To study the effects of variations in mixing, we slowly altered the mixing algorithm in \cite{2012ApJ...749...91F}.  By varying the growth time for the convective instabilities, we alter the energy injection and energy region.  Studying variations in these properties, a wide range of explosion energies and properties can be produced for the same basic progenitor~\citep{2018ApJ...856...63F}.  This study found that a wide range of explosion properties and remnant masses could be produced by varying the energy injection rate and region.

The final fate of these massive stars depends both on the growth time of the convection and the nature of the stellar core.  \cite{2020ApJ...890..127C} found that, for the progenitor models in their study, they could produce an explosion with a sufficiently high mixing length value.  Increasing the mixing length accelerates the convective growth and, as this convection grows, the more likely it is to explode.  This result demonstrates the same trends seen in other studies altering convection~\citep[e.g.][]{2018ApJ...856...63F} which form the basis for our models.  
In this section, we use these trends from 1-dimensional explosion models, integrating them into prescriptions that incorporate the stellar structure, to determine the remnant mass for a given progenitor.  

\begin{figure}
\plotone{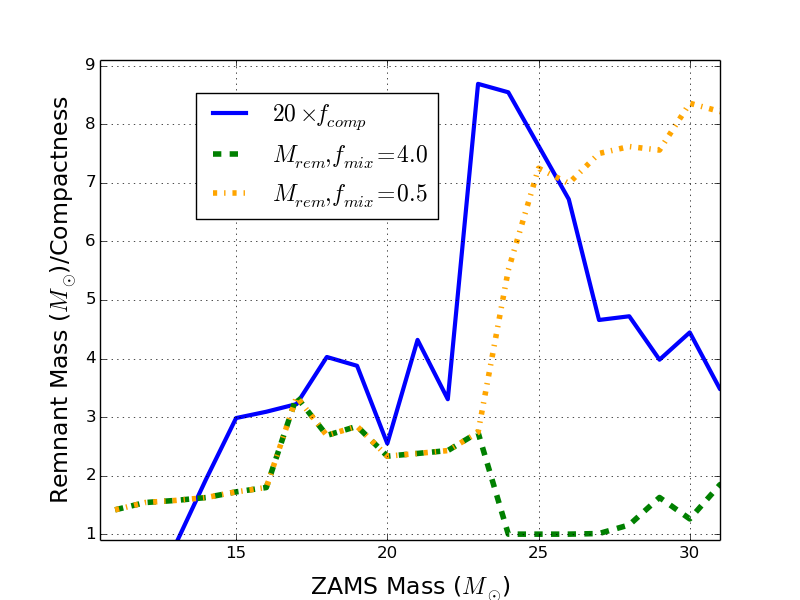}
\caption{Compactness parameter (determined at $2.5\,M_\odot$) and remnant mass predictions based on the model in ~\cite{2001ApJ...554..548F} using the solar metallicity models from~\cite{2002RvMP...74.1015W} as a function of the zero-age main-sequence (ZAMS) mass.  Some of the characteristics of the compactness parameter mimic each other.  For example, as the compactness parameter rises with incrasing ZAMS mass, so do the remnant masses in our model.  As the compactness parameter drops beyond this initial rise, so too does the remnant mass.  But the subsequent evolution of the remnant masses depends upon our assumption of the convective growth time.} 
\label{fig:fcomp}
\end{figure}

The stellar structure at collapse can play an even more critical role than convective growth time in determining the stellar fate.  Although it is well known that a single compactness parameter is insufficient to determine the final fate of a  model~\citep[e.g.][]{2020MNRAS.491.2715B}, our models are based on a prescription that follows the time-evolution of the infalling stellar material~\citep{2001ApJ...554..548F,2012ApJ...749...91F}.  To compare this method to approaches solely using compactness parameter, we plot in Figure~\ref{fig:fcomp} both the compactness parameter and our final remnant masses (using two different convective growth times).  Although there are trends between remnant masses from our prescription and the compactness parameter, as with the models of \cite{2020ApJ...890..127C,2018ApJ...856...63F}, the ultimately fate is more complex than this.

The final structure of the star, which dictates the fate of the collapsing star, whether using a simple compactness parameter or our more complex prescription, can vary dramatically with stellar model~\citep{2018ApJ...860...93S}.  The multiple carbon and oxygen shell-burning phases at the end of a star's life can evolve very differently depending on small amounts of mixing.  The largest variation in this evolution occurs in stars with zero-age main sequence mass lying between 20 and 30\,M$_\odot$.  These variations can alter the fate, allowing supernova explosions in these more massive stars.   \cite{2012ApJ...749...91F} used a single set of stellar models~\citep{2002RvMP...74.1015W} to guide their stellar fates.  Understanding this convection is critical to understanding the fate of stars in this 20-30\,M$_\odot$.  We will discuss this further below in our parameterized models.

\begin{figure}
\plotone{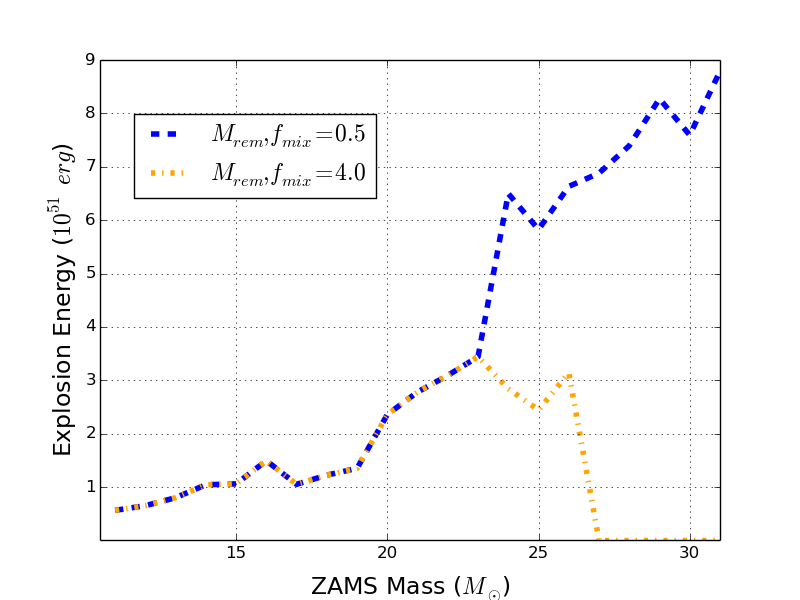}
\caption{Explosion energies based from the models producing the remnant masses in Figure~\ref{fig:mrem}.  For low-mass stars (with small compactness parameters - below about 14\,M$_\odot$), the explosion energy is low (below $10^{51}{\rm \, erg}$).  These progenitors also produce low-mass remnants (Fig.~\ref{fig:mrem}).  For massive stars above 23\,M$_\odot$ to produce neutron star remnants, the explosion energy must be very strong.} 
\label{fig:fene}
\end{figure}

Final explosion energies are more difficult to calculate because 1-dimensional models typically don't include additional energy from further fallback~\citep[although see][]{2018ApJ...856...63F}.  However, our models do calculate the energy in the convective region when it is able to blow off the infalling material and this provides a rough estimate of the explosion energy (explosion energy is lost in the ejection of the star but energy is gained through fallback, so this value can be higher or lower than the final explosion energy).  The corresponding explosion energy for these zero-age main-sequence progenitors is shown in Figure~\ref{fig:fene}.

\begin{figure}
\plotone{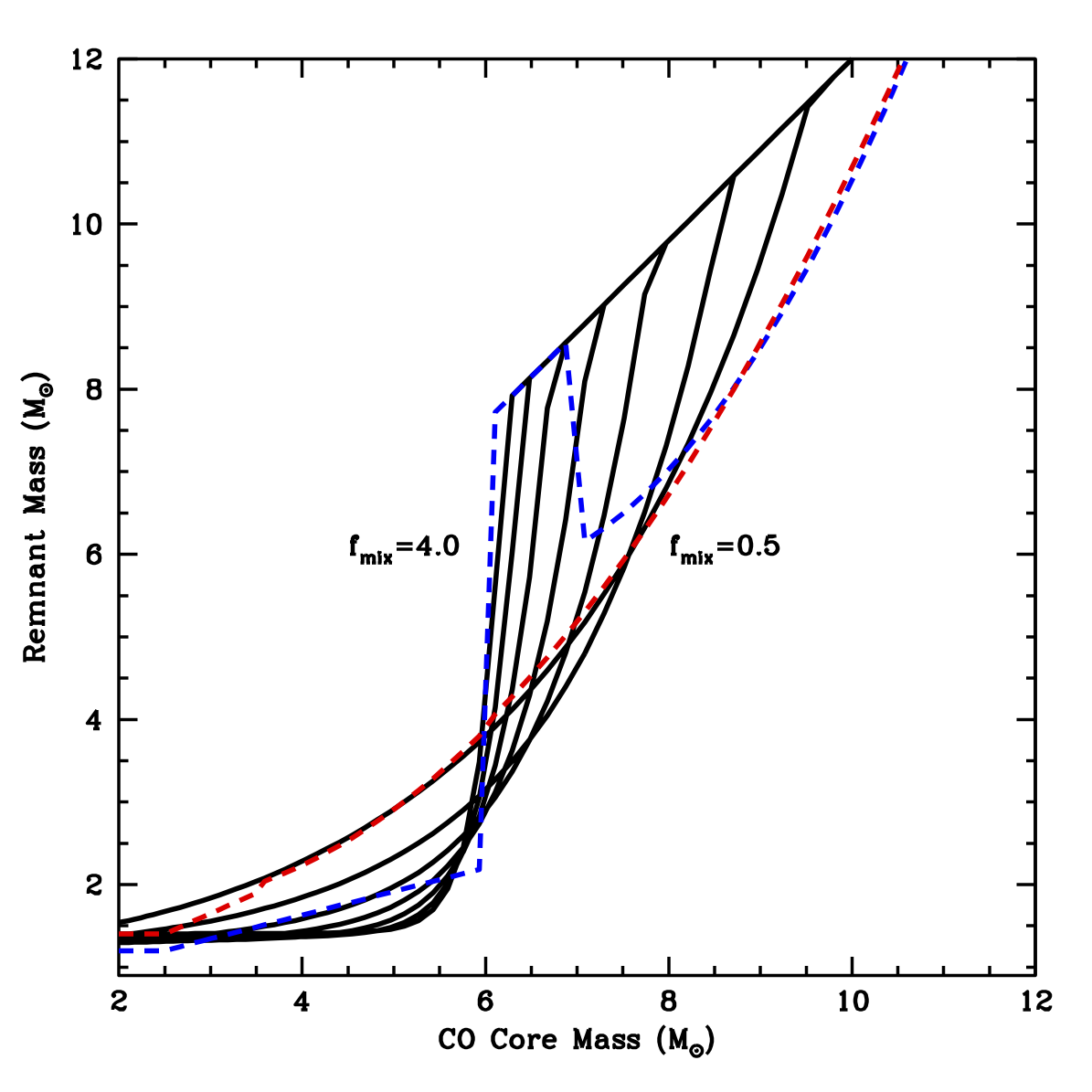}
\caption{Baryonic remnant mass as a function of CO core mass for our new parameterized models (black curves) compare to the prescriptions of \cite{2012ApJ...749...91F} (blue and red dashed curves).  For our new parameterized models, we use $M_{\rm crit}=5.75$ and vary $f_{\rm mix}$ from 0.5 to 4.0.} 
\label{fig:mrem}
\end{figure}

If we assume that there is a smooth trend in core properties with respect to CO core mass, we can develop a simple smooth fit by varying the mixing efficiency set by the growth time.  A set of models fill in the solutions from \cite{2012ApJ...749...91F} for the final baryonic remnant mass ($M_{\rm remnant}$) is given by:
\begin{align}
    M_{\rm remnant} = & 1.2+0.05f_{\rm mix}+0.01 (M_{\rm CO}/f_{\rm mix})^2 + \nonumber \\ 
    & e^{f_{\rm mix}(M_{\rm CO}-M_{\rm crit})}
    \label{eq:mrem}
\end{align}
where $M_{\rm CO}$ is the Carbon/Oxygen core mass, $M_{\rm crit}$ is a critical mass for black hole formation (for our models, we typically use $M_{\rm crit} = 5.75\,M_\odot$) and $f_{\rm mix}$ is a parameter describing the mixing growth time. This mass can not exceed the total stellar mass at collapse so the final remnant remnant mass (baryonic) is:
\begin{equation}
    M_{\rm remnant} = min(M^{eq.~\ref{eq:mrem}},M_{\rm collapse})
\label{eq:rem}
\end{equation}
where $M^{eq.~\ref{eq:mrem}}$ is the remnant mass from equation~\ref{eq:mrem},  $M_{\rm collapse}$ is mass of the star at collapse.  The resultant remnant masses using this prescription are overlaid upon our calculated remnant masses in figure~\ref{fig:mrem} where we have varied $f_{\rm mix}$ from 0.5 to 4.0 where $f_{\rm mix}$=4.0 corresponds to a fast growth time (rapid model in \cite{2012ApJ...749...91F}; blue line in Fig.~\ref{fig:mrem}) and $f_{\rm mix}=0.5$ corresponds to a slow growth time (delayed model in \cite{2012ApJ...749...91F}; red line in Fig.~\ref{fig:mrem}).

We use CO core mass as the principle stellar-structure quantity for the fate of these stars because most population synthesis models include its evolution with binary mass transfer and mass-loss.  However, the CO core mass does not capture the full structure if the core that scientists characterize by either the compactness parameter of the more detailed infall prescription used in our models.  In addition, the differences between the CO mass and this structure also depend on metallicity.  Barring population synthesis models that include the evolution of these more accurate prescriptions, we are currently limited to simplified remnant descriptions depending on properties like the CO core mass.  An alternate approach could use the ZAMS mass alone, but this would not include evolution caused by binary mass transfer and loss.

As we discussed above, late-stage carbon- and oxygen-shell burning stages depend sensitively on the progenitors and progenitor models.  In general, the remnant mass increases (as in our equation~\ref{eq:mrem}) with CO mass.  But this trend can turn around briefly for stars with final CO cores in the $6-9\,M_\odot$ mass range.  In \cite{2012ApJ...749...91F}, final CO cores above $6M_\odot$ formed black holes and, although the details of the convection allowed some stars with CO core masses above $6M_\odot$ to produce supernova explosions, the explosions were too weak to produce neutron stars.  In their models, the final black hole 
masses were below the mass expected from a direct collapse (no explosion), limiting the final black hole mass in a region roughly between $6.5-10\,M_\odot$ to roughly 75\% of the mass of the star at collapse.  One possible parameterization of this effect is to limit the mass of the black hole to a fraction ($f_{\rm CO burning}$) of the collapse mass between a range $M_{\rm low}$ and 
$M_{\rm high}$ where, for \cite{2012ApJ...749...91F}, these values are roughly:  $f_{\rm CO burning}=0.75$, $M_{\rm low}=6.5$, and 
$M_{\rm high}=10.0$.  We do not include this effect in the populations discussed in the rest of this paper.

\begin{figure}
\plotone{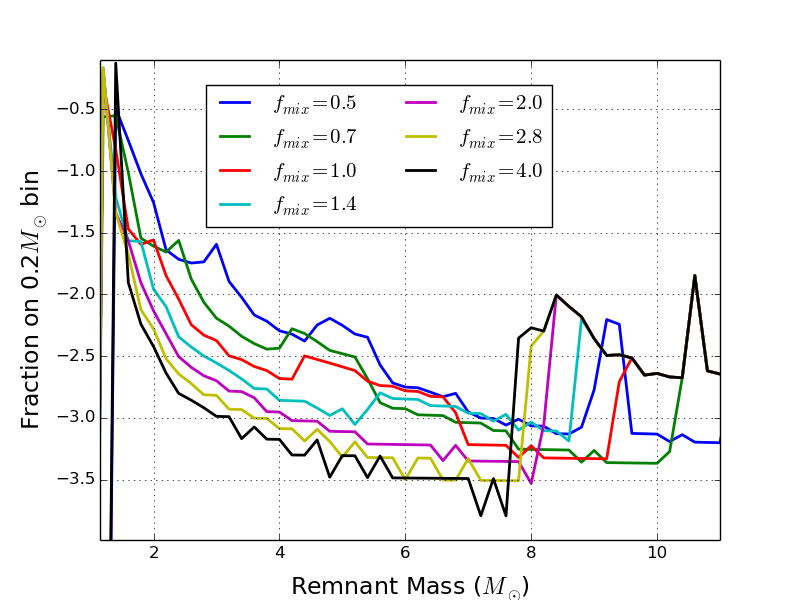}
\plotone{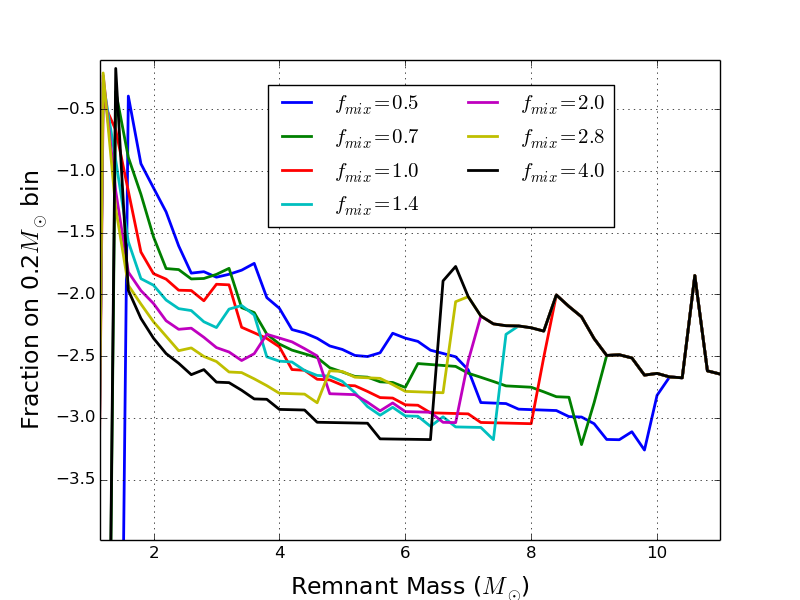}
\caption{Distribution of baryonic compact remnant masses for seven different values of $f_{\rm mix}$.  The depth and width (in remnant) space depends upon this mixing growth rate.  The position in mass coordinates depends on $M_{\rm crit}$:  the top panel uses a value of 5.75\,M$_\odot$, the bottom panel uses a value of 4.75\,M$_\odot$.  None of these models predict no compact remnant masses in the mass gap, but the fraction can drop by a factor of 10 between slow and fast growth times.  } 
\label{fig:imf}
\end{figure}

If we assume stars follow the Salpeter ($\alpha_{\rm IMF}=2.35$) initial mass function (IMF) and the CO and He cores from Kepler models, the prevalence of a gap in the compact remnant masses depends upon this mixing factor.  Figure~\ref{fig:imf} shows the distribution using 7 values of $f_{\rm mix}$:  0.5, 0.7, 1.0, 1.4, 2.0, 2.8 and 4.  The top panel uses $M_{\rm crit}=5.75\,M_\odot$, the bottom panel uses $M_{\rm crit}=4.75\,M_\odot$.  For the rest of the figures in this paper, we use $M_{\rm crit}=5.75\,M_\odot$ as this value provides a closer match to past studies~\citep{2012ApJ...749...91F}.  However, this value can change depending on physical factors in the supernova engine (e.g. neutrino physics, equation of state).  The larger mixing factors produce larger and deeper gaps in the remnant mass distributions which can be compared to observations to constrain the growth of convection in the engine.  Although all of these models produce compact remnant masses, the number of systems in the gap can vary by a factor of 10.  

This simple model produces a smooth trend in the remnant masses with progenitor mass.  Variations in the remnant mass distributions arise from variations in the CO core mass in our coarse-sampled (in mass) stellar models. But the variations in stellar mixing produce even stronger variations in the CO core mass and, combined with variations in seeds of convection in the supernova engine, much larger variations could exist in the mass distribution.  To study the the full distribution, we need to include this stochastic variation.


\subsection{Stochasticity from Convection in the Star and in the Supernova Engine}

The simple smooth remnant mass/CO core mass distribution does not account for variation in the structure and uniformity of the star at collapse caused by stellar mixing.  Here we introduce two modifications to our smooth profile to capture the stochasticity that is currently known to be present in supernova explosion calculations.  The first captures the growth time variation that can be caused by asphericities in the collapsing core that then seed the convection.  The latter captures the dramatic structural differences seen in some stars where, for example, the core of a $14.9\,M_\odot$ star follows our basic trend producing a NS, but the core of a $15.0\,M_\odot$ collapses to form a black hole.  

\begin{figure}
\plotone{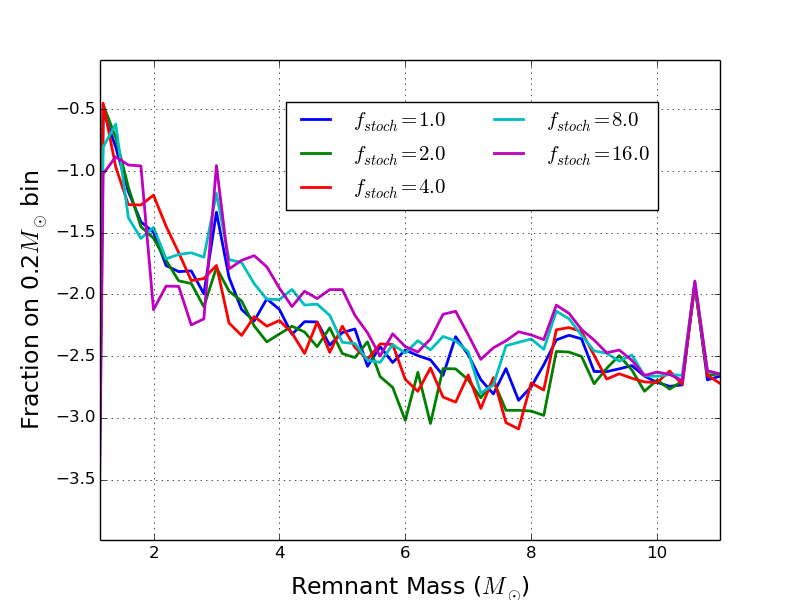}
\caption{Distribution of baryonic compact remnant masses for 5 different values of $f_{\rm stoch}$ and 10 million sampling points.  The peak near 3\,M$_\odot$ occurs for systems produced with very low mixing growth times.} 
\label{fig:imfs1}
\end{figure}

Asymmetries in the collapsing core provide the seeds for convection that then change the growth time of the convection.  To capture this effect, we sample $f_{\rm mix}$ from a Gaussian:
\begin{equation}
    F(f_{\rm mix}) = f_{\rm norm} e^{-f_{\rm stoch}(f_0-f_{\rm mix})^2}
\end{equation}
where $F(f_{\rm mix})$ is the fraction of stars with a value of the mixing parameter equal to $f_{\rm mix}$, $f_0$ is the average value of $f_{\rm mix}$, $f_{\rm stoch}$ denotes the variation in the growth times, and $f_{\rm norm}$ is a normalization factor.  Using $\alpha_{\rm IMF}=2.35$ and our Kepler models, the remnant mass distribution for $f_0=1$, $f_{\rm stoch}=1, 2, 4, 8, 16$ produces remnant mass distributions in Figure~\ref{fig:imfs1}.  For this distribution, we use a Monte Carlo sampling of this Gaussian distribution using 10 million sampling points.  By using a low normalization factor, we have skewed these distributions toward low values of mixing.  

Broader changes are required to capture the more dramatic variation seen in calculations where slight changes during a shell burning episode dramatically alter the size of a burning layer, ultimately altering the structure of the core at collapse and its final fate.  One way to implement these changes is to use the detailed grids with stochastic results to determine CO core masses instead of using a formula fit to the CO core mass that is smooth.  The evolutionary changes that alter these fates are typically reflected in the CO core mass.  However, because different codes (and indeed different initial grid resolutions) can produce very different results (the nature of a stochastic process whether real or numerical), no grid of models represents the ultimate answer.  It is not even obvious how to interpolate between masses in such stochastic simulations.  

\begin{figure}
\plotone{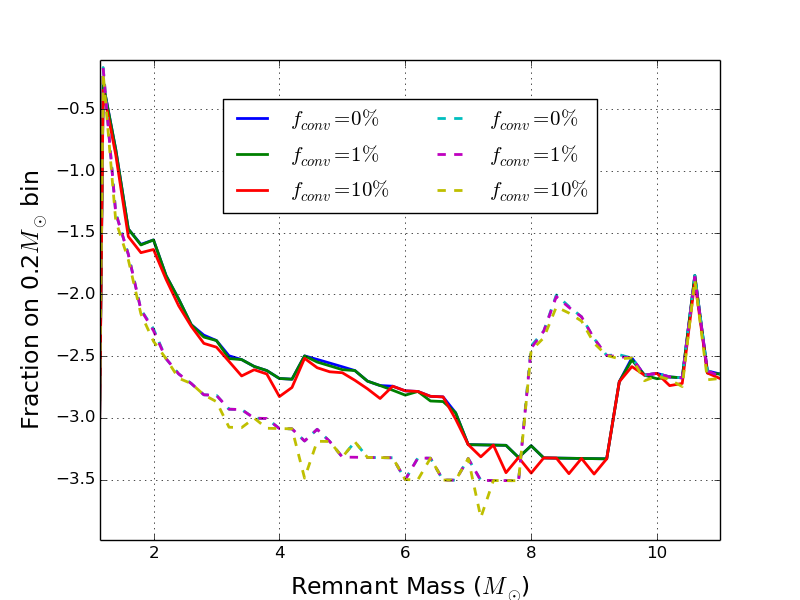}
\caption{Distribution of baryonic compact remnant masses for 5 different values of $f_{\rm conv}$ and 10 million sampling points. The solid curves correspond to $f_{\rm mix}=1$ and the dashed curves correspond to $f_{\rm mix}=8$.  The effect of the variation in the fraction of randomly sampled systems does not dramatically change the integrated mass distribution.  The $f_{\rm conv}=0,1\%$ are nearly identical and are difficult to distinguish in population studies.} 
\label{fig:imfs2}
\end{figure}

We instead propose a more parameterized approach where we assume some percentage $f_{\rm conv}$ of stars have a very different stellar burning phase where the fate ranges from the proto-compact object and the mass of the star at collapse.  The proto-compact object mass is $M_{\rm proto}$ from equation 18 of \cite{2012ApJ...749...91F}.  By setting $f_{\rm conv}$ to 1-10\%, we alter the remnant mass of 1-10\% of our CO cores to a mass sampled from our entire set of systems focusing on stars with $15<M_{\rm ZAMS}/M_\odot<50$ (ZAMS corresponds to the Zero-Age Main Sequence mass).  For an individual system, this means that some subset of stars can produce a black hole instead of a neutron star (or vice versa).  Using $\alpha_{\rm IMF}=2.35$ and our Kepler models, the remnant mass distribution for $f_{\rm conv}=0,1,10\%$ (again using 10 million sampling points) and  $f_{\rm mix}=1,8$ produces remnant mass distributions in Figure~\ref{fig:imfs2}.  Here we have used $f_{\rm mix}=8$ to study the extreme solutions.  In general, it is likely that $f_{\rm mix}$ lies between 0.5-4.  Although this stochasticity can dramatically change the remnant mass of a given CO core, this variation does not affect the integrated remnant mass distribution if the sample size is large.

\begin{figure}
\plotone{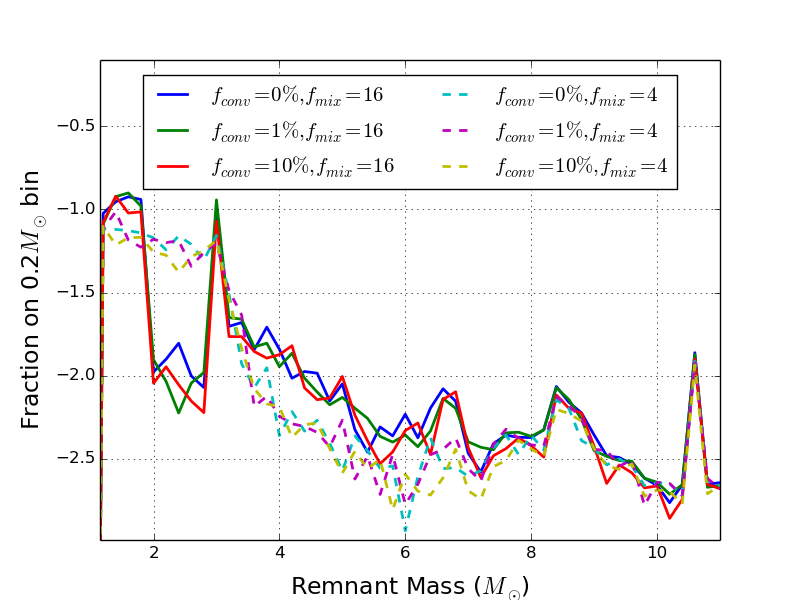}
\caption{Remnant mass distributions including stochasticity effects from both the progenitor convection that affects the stellar structure and instabilities ($f_{\rm conv}$) that dictate the growth of the convection in the supernova engine ($f_{\rm stoch}=4$ for two values of $f_{\rm mix}$).  The effects of stochasticity are much smaller than the value for mixing ($f_{\rm mix}$).  In this figure $f_0$ is set to 1.} 
\label{fig:imfs3}
\end{figure}

As we show in Figure~\ref{fig:imfs2}, stochasticity in the progenitor is unlikely to affect the remnant mass distribution if the sample size of remnants is large (e.g. for population synthesis calculations studying a large number of binaries) because such large sampling smooths out these stochastic variations.  However, the current observed distribution of compact remnants is small.  Stochastic variations can explain single observations of peculiar observations:  e.g. an observed system with a black hole forming from a 15\,M$_\odot$ progenitor or a neutron star produced by a 30\,M$_\odot$ progenitor.  Population studies using small sampling sizes could consider including stochasticity.  Figure~\ref{fig:imfs3} shows the remnant distribution with stochastic model using a sampling of only 1,000 points and combining both progenitor and engine stochasticities discussed in this section.  Although varying the amount of stochasticity in the progenitor ($f_{\rm conv}=0,1,10\%$), the effect of this stochasticity remains minimal on the remnant distribution.  From this analysis, unless a study is trying to explain the masses of a small number of systems or population studies are using samplings below 1,000 systems, stochasticity is not an important effect.

\section{Angular Momentum and Remnant Masses}
\label{sec:rotation}

Two primary black hole formation scenarios have been identified~\citep{1999ApJ...522..413F}:  fallback and direct collapse formation.  Most studies of compact remnant formation focus on determining the dividing lines between different compact remnant fates (NS, fallback BH, direct BH) and the amount of mass accreted in the fallback BH formation. It is generally assumed that in the direct-collapse BH formation that all of the mass collapses to form a black hole.  This is not the case for rapidly rotating stars.  Since the first models of collapsar jets through a massive star, it has been known that the jet drives a shock that propagates around the star, ejecting much of the stellar mass~\citep{1999ApJ...524..262M,2001ApJ...550..410M}.  This jet-driven mass ejection limits the final mass of the black hole.  For progenitor stars with sufficient angular momentum to form a disk that drives a jet, this mass ejection process can eject much of the star even if the star collapses directly to a black hole.  To determine the mass ejection from disk- and jet-driven outflows, we must understand disk formation and the angular momentum in stars.  

To form a disk, the star must have sufficient angular momentum such that infalling material of the imploding star is centrifugally supported in a disk.  In general, the specific angular momentum of a star increases with radius.  For example, if the star were strictly coupled with a constant angular velocity ($\omega$), the angular momentum of the star increases with the square of the radial position in the star ($j = I \omega \propto \omega r^2$).  Many stellar evolution models assume only mild coupling in the stars, allowing the inner core to spin more rapidly than the outer layers.  But even for these stars, the angular momentum of a star increases as we move outward in mass coordinate~\citep[see, for example][]{2000ApJ...528..368H}.  The imploding star continues to accrete onto its core until the angular momentum of infalling stellar material is sufficiently high to produce a disk.  Because the angular momentum increases with radius, even if the innermost material doesn't have enough angular momentum to form a disk, the outer layers of the star may.  The innermost material will accrete onto the black hole, increasing its mass until the higher-angular momentum material forms a disk\footnote{It is for this reason that scientists argued that collapsar gamma-ray bursts were produced by black hole accretion disks~\citep{1993ApJ...405..273W}.  In contrast, neutron star accretion disks or rapidly-spinning NS systems powering $10^{51}\,{\rm erg}$ bursts is expected to be extremely rare~\citep{2019EPJA...55..132F}.}.  

\begin{figure}
\plotone{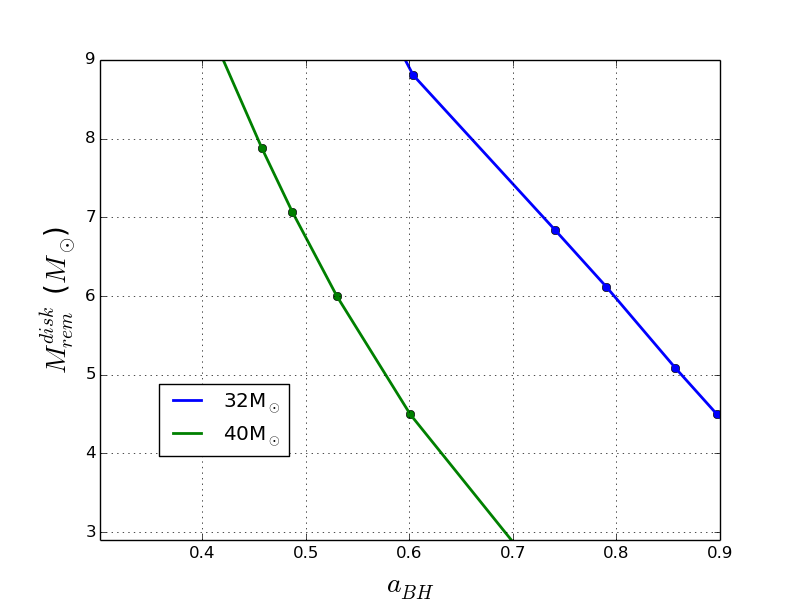}
\caption{Remnant mass at the time of disk formation as a function of the black hole mass spin period using two stars produced with the GENEC stellar evolution code~\citep{2000A&A...361..159M} assuming weak coupling between burning layers.  We then artificially raise and lower the angular momentum of these stars to produce a range of spin periods using their stellar profile.} 
\label{fig:mdisk}
\end{figure}

We can use the angular momentum profile to determine the mass of the core when a disk forms.  This angular momentum also sets the spin of the black hole produced when the disk forms.  Figure~\ref{fig:mdisk} shows the mass of the remnant at the time of the formation of a 300\,km disk as a function of the final black hole mass spin period:  $a_{\rm BH} = (cJ_{\rm BH})/(GM_{\rm BH}^2)$ where $c$ is the speed of light, $J_{\rm BH}$ is the black hole angular momentum, $G$ is the gravitational constant and $M_{\rm BH}$ is the black hole mass.  This figure uses two stars from the GENEC stellar evolution code~\citep{2000A&A...361..159M} with weak coupling where we artificially raise and lower the spin period to produce a range of angular momentum profiles. The faster the spin rate, the earlier a disk forms and the higher the value of the final black hole spin.  After the disk forms, the core will continue to accrete until a disk-driven jet is able to eject the star.  In this manner, the mass of the remnant mass at disk formation in Figure~\ref{fig:mdisk} shows the minimum remnant mass of a collapsing core as a function of its angular momentum.  If the coupling is strong, as suggested by \cite[e.g.][]{2019ApJ...881L...1F,2019MNRAS.485.3661F}, the angular momentum in the helium core is insufficient to form a disk.

\begin{figure}
\plotone{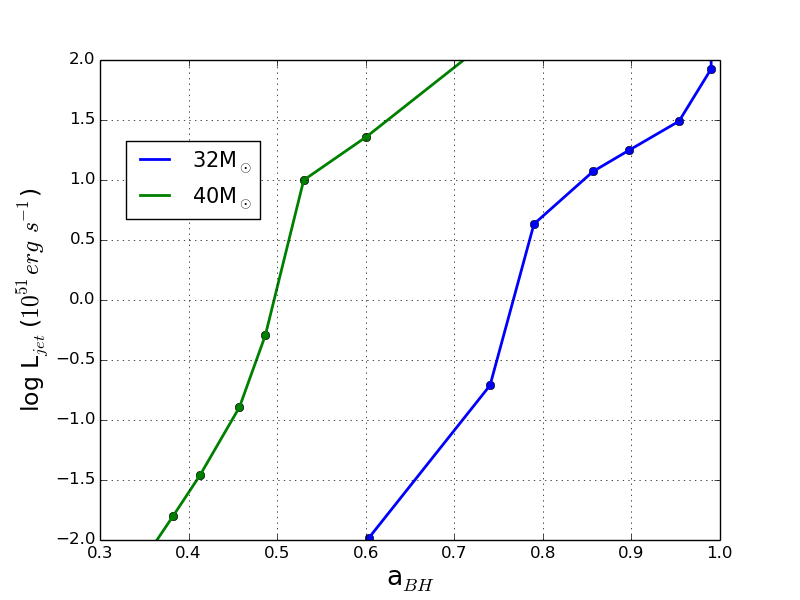}
\caption{Energy of the jet produced as a function of black hole spin based on the structure of 2 progenitor stars and varying the spin of these stars after the giant phase (as in Figure~\ref{fig:mdisk}).  The angular momentum of the stars sets the mass of the remnant when the disk forms (Figure~\ref{fig:mdisk}) and the spin of the black hole.  This plot assume the energy is then derived from equation~\ref{eq:lgrb}.} 
\label{fig:lgrb}
\end{figure}

This black hole accretion disk is the power source invoked in the collapsar gamma-ray burst engine.  We will assume that the efficiency at which the disk ejects mass is proportional to the energy expected in this black hole accretion disk engine.  There have been many studies of these black hole accretion disks and the energy of their outflows.  For our study, we use a simple formula derived by \cite{2003ApJ...591..288H} based on the disk models from \cite{1999ApJ...518..356P} using the \cite{1977MNRAS.179..433B} energy mechanism:
\begin{align}
L_{\rm BZ} =  & 10^{50} a_{\rm spin}^2 10^{0.1/(1-a_{\rm spin})-0.1} \nonumber \\
& \left( \frac{M_{\rm BH}}{3 M_\odot} \right)^{-3} \frac{dM/dt}{0.1M_\odot/s} {\rm \, erg \, s^{-1}}
\label{eq:lgrb}
\end{align}
where $dM/dt$ is the accretion rate\footnote{The canonical long-duration gamma-ray burst has powers of $10^{50}-10^{51} {\rm \, erg s^{-1}}$}.  Figure~\ref{fig:lgrb} shows this energy as a function of the final black hole spin using the same models as Figure~\ref{fig:mdisk}.  

Even with these approximations, it is still not straightforward to determine the final black hole mass because we must estimate the amount of disk material that accretes while its outflow and jet ejects the rest of the star.  For powerful jet-driven explosions (high accretion rates), the outflow can eject much of the infalling star.  In such an extreme, we assume that the accretion of 1\,M$_\odot$ of material is sufficient to eject the entire star.  If this occurs, instead of our standard black hole remnant mass formula (for direct collapse, this value is close to the mass at collapse), the remnant mass for a rotating system $M^{\rm rot}_{\rm BH}$ can be as low as: 
\begin{equation}
    M^{\rm rot}_{\rm BH}=max(3.0,7.5-15.0(a_{\rm BH}-a_{\rm crit})) M_\odot 
\end{equation}
for $a_{\rm BH}>a_{\rm crit}$ where $a_{\rm crit}$ is set by the models ($\sim$ 0.7,0.45 for the 32,40\,M$_\odot$ stars in our study) and $a_{\rm BH}$ is the final black hole spin.  If $a_{\rm BH}<a_{\rm crit}$, we assume the angular momentum in the star is insufficient to produce a disk that produces a sufficiently energetic outflow to alter our standard remnant mass.  

\begin{figure}
\plotone{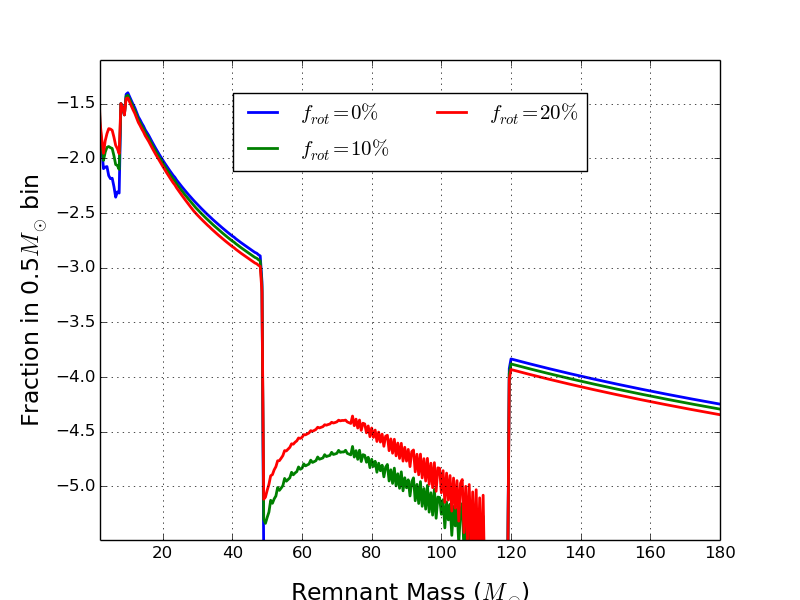}
\caption{Distribution of black hole remnant masses assuming different fractions of systems with sufficient angular momentum to produce a disk:  0, 10, and 20\%.} 
\label{fig:imfbh}
\end{figure}

Unfortunately, we do not know the angular momenta of massive stars prior to collapse.  Stars are born rotating and, if the different burning layers do not couple, the core could be spinning rapidly at collapse.  A weak coupling (typically assumed to be produced by magnetic fields), has been argued as an explanation of the spin periods behind pulsars.  If we assume a set fraction of our black hole forming stars have a spin value that is sufficiently high to cause the collapsed core to eject half of its final mass, we can use our IMF to predict BH mass distributions for our 2 different mass-limit extremes (Figure~\ref{fig:imfbh}).  This figure varies the number of high-rotating systems from 0 to 20\% of the black hole forming stars.

For stars that lie above the pair-instability mass gap corresponding to a ZAMS mass above $\sim 260 {\rm \, M}_\odot$, it is possible that, unlike BH-forming scenarios where first a Proto-NS is formed that accretes until it collapses to a BH (first forming a low mass BH), the star can first form a proto-BH~\citep{2001ApJ...550..372F}.  In such a case, the remnant mass will be closer to the proto-black hole mass, producing gravitational masses above 60-70\,M$_\odot$.  We add the distribution from a model assuming proto-BH formation in our plot of black hole masses (Figure~\ref{fig:imfbh}). 

\section{Binary Spin Up and Remnant Masses}
\label{sec:binarybh}

Although measurements of X-ray binaries suggests that many collapsing stars have high angular momenta, these measurements are both difficult to make and limited to a select set of formation scenarios requiring short-period, accreting binaries~\citep{2015PhR...548....1M,2021arXiv211109401B}.  Gravitational wave measurements of black hole binaries suggest that many black hole systems have low spin values~\citep{2019ApJ...882L..24A}, suggesting that angular momentum coupling in massive stars is strong~\citep[e.g.][]{2019ApJ...881L...1F,2019MNRAS.485.3661F}.  If this coupling is strong, forming fast-spinning black holes and collapsed cores with disks will be limited to systems that can be spun up prior to collapse.  One method to spin up these stars is through tidal locking~\citep{2016MNRAS.462..844K,2017MNRAS.467.2146K,2017ApJ...842..111H,2018A&A...616A..28Q,2018MNRAS.473.4174Z}.  Incorporating a solution only using this tidal-locking spin-up, \cite{2020A&A...636A.104B,2021ApJ...921L...2O} were able to use binary population models to match the observed $\chi_{\rm eff}$ values observed by gravitational-wave observatories~\citep{2019ApJ...882L..24A}.

In this section, we study the role of this tidal-locked angular momentum model using the results of binary population synthesis models.  We use the population synthesis code {\tt StarTrack}~\citep{2008ApJS..174..223B} to calculate evolution of four populations of massive binary stars. Each population consists of $10,000$  massive binary systems, in which initial primary mass is drawn from power-law IMF with exponent $\alpha=-2.3$ and in the ZAMS mass range $M_{\rm ZAMSA}=5-200M_\odot$, and secondary initial mass is in sampled from a ZAMS range $M_{\rm ZAMSB}=3-200M_\odot$ and is obtained from uniform mass ratio ($q=M_{\rm ZAMSB}/M_{\rm ZAMSB}$) distribution. For each population we study two  metallicities:  $Z=0.1Z_\odot,\ 0.01Z_\odot$ (with $Z_\odot=0.014$) and two mixing parameters:  $f_{\rm mix}=0.5,\ 4.0$. These massive binaries are potential progenitors of double compact objects (NS-NS, BH-NS and BH-BH systems). Note that formation of double compact objects is expected only from a very small fraction of massive binaries.

In our calculations we assume standard wind losses for massive stars: O/B star winds~\citep{Vink2001}, LBV mass loss, and WR star winds \citep[specific prescriptions for these winds are listed in Sec.~2.2 of][]{Belczynski2010b}.  We treat the accretion onto compact objects during the Roche lobe overflow and from stellar winds using the analytic approximations presented by \cite{King2001} and by \cite{Mondal2020}, and limit accretion during the common envelope (CE) phase to $5\%$ of the Bondi rate~\citep{MacLeod2017}. In calculations of remnant mass, we employ the following scheme:  for NSs and low-mass BHs we employ the prescriptions presented in this study where we calculate the BH/NS mass with $f_{\rm mix}=0.5,\ 4.0$ but do not account for effects of stochasticity. 
We allow for formation of BHs up to $90\,M_\odot$~\citep[as detailed in][]{2020ApJ...905L..15B} taking into account recent advancements in understanding effects of pair-instabilities on evolution of massive stars~\citep{Farmer2020,Costa2021,Farrell2021}. We do not produce black holes from systems above the pair-instability limit (stars above $\sim 260\,M_\odot$).  We employ 
the fallback decreased NS/BH natal kicks with $\sigma=265{\rm \,km\,s^{-1}}$, we do not allow CE survival for Hertzsprung gap donors (submodels B in our past calculations). Note that we allow for standard CE development criteria and most double compact objects form through CE channel~\citep{Belczynski2016b}, and not through stable Roche lobe overflow channel~\citep{2017MNRAS.471.4256V,2021A&A...651A.100O}.  The most updated description of {\tt StarTrack} is given by \cite{2020A&A...636A.104B} and the model M30 in this study describes our standard choices of input physics. 

\begin{figure}
\includegraphics[scale=0.32,angle=0]{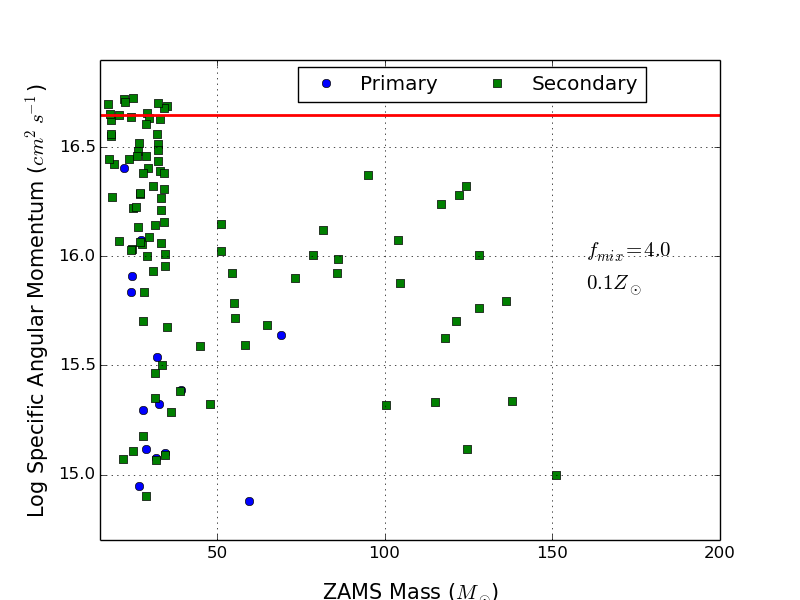}
\includegraphics[scale=0.32,angle=0]{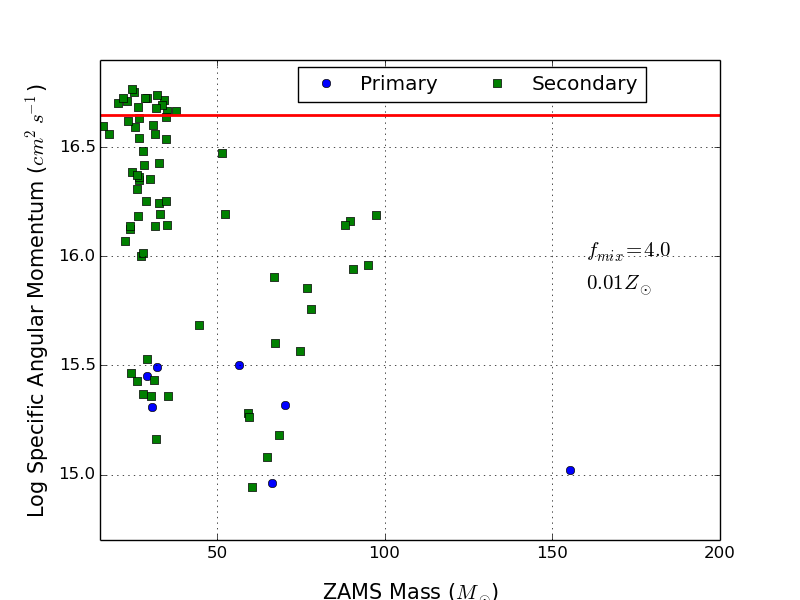}
\includegraphics[scale=0.32,angle=0]{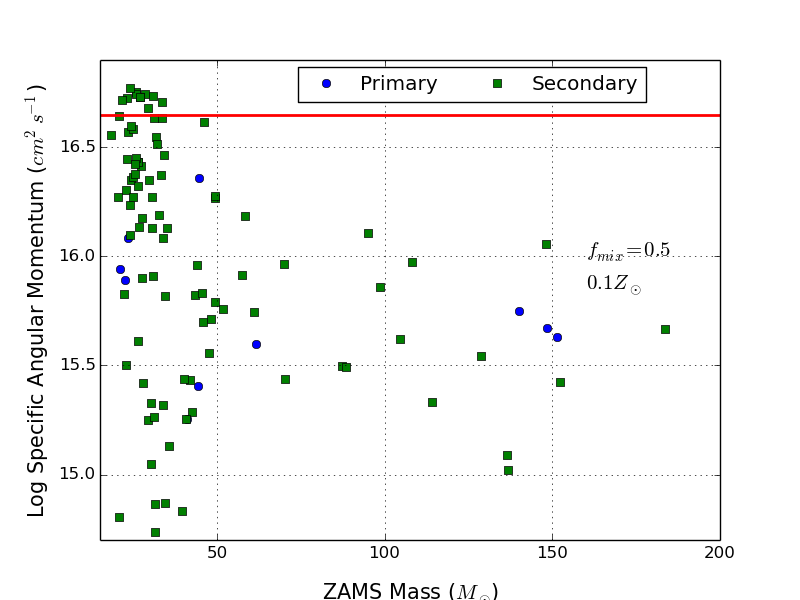}
\includegraphics[scale=0.32,angle=0]{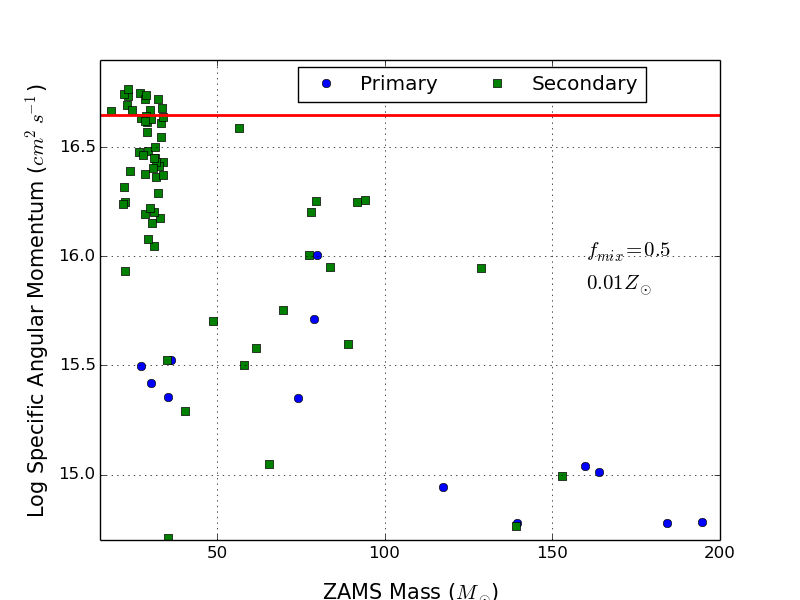}
\caption{Log specific angular momentum at $10^{10}$\,cm assuming tidal-locked binaries from our binary population synthesis calculations.  The 4 panels show the {\tt StarTrack} results at two metallicities (0.01,0.1\,Z$_\odot$) and two mixing parameters ($f_{\rm mix}=0.5,4.0$).  In the binary, the primary has the higher ZAMS mass and the secondary has the lower ZAMS mass.  The horizontal line corresponds to the angular momentum required to make a 50\,km disk around a 3\,M$_\odot$ BH.} 
\label{fig:angmr10}
\end{figure}

\begin{figure}
\plotone{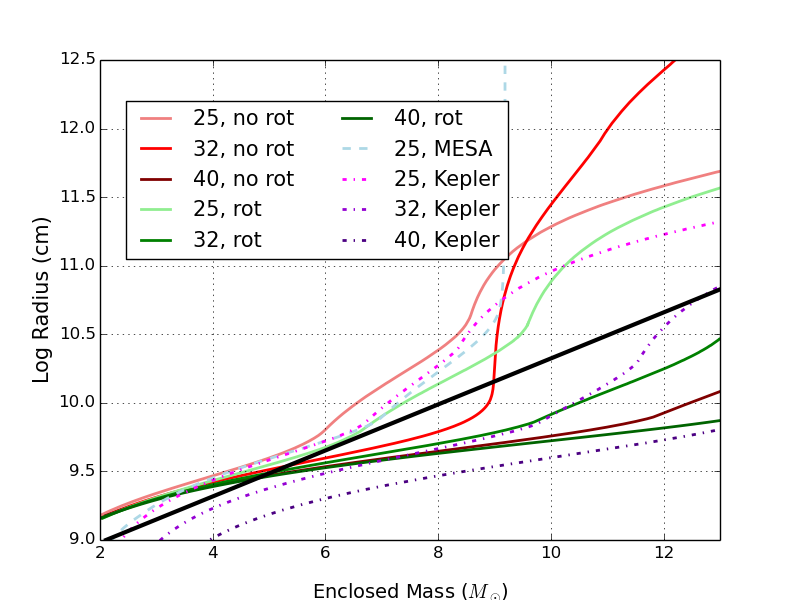}
\caption{Log radius versus enclosed models for 3 different stellar masses:  25,32, and 40\,M$_\odot$ using 3 different stellar evolution codes:  KEPLER, GENEC, and MESA.  The GENEC and MESA models are from~\cite{2020A&A...636A.104B}.  The KEPLER models are from ~\cite{2002RvMP...74.1015W}.  The GENEC calculations include both rotating and non-rotating models.  However, the differences in this internal structure between stellar evolution codes is far greater than the differences between progenitor properties.  The solid black line shows an approximate fit to these models.} 
\label{fig:rvmf}
\end{figure}

From the progenitors of BH-BH binaries from these calculations, we obtain orbital properties prior to the collapse of each star.  These systems undergo binary interactions that tighten the orbits and eject mass.  To determine the spin of these stars, we assume direct tidal locking of the star at collapse for systems with periods less than 1.3\,d (assuming the timescale for tidal locking is longer than the evolution timescale for binaries with greater orbital periods as did~\cite{2020A&A...636A.104B}).  For each of the binaries in our population synthesis model, we calculate the orbital period prior to the collapse to a black hole of both the primary and the secondary star.  For these calculations, we refer to the primary star in the system as the star that has a larger ZAMS mass in the binary (the secondary star has the lower ZAMS mass in the system).  The specific angular momenta at a radius of $10^{10}\,{\rm cm}$ in each of these stars is shown in Figure~\ref{fig:angmr10}.  In this figure, we compare results from binary population synthesis models using two metallicities and two mixing parameters.  We use $10^{10}\,{\rm cm}$ because, for the massive stars in this study, this lies well within the helium layer and provides a maximum extent (highest angular momentum) for systems whose compact remnant is not so massive that the formation of the disk will still drive powerful jets.  For the material to hang up in a disk, the centrifugal acceleration must exceed that of gravity above the innermost stable circular orbit (given for a non-rotating black hole):  
\begin{equation}
    r_{\rm ISCO} = 6 G M_{\rm BH}/c^2
\end{equation}
where $G$ is the gravitational constant, $c$ is the speed of light, and $M_{\rm BH}$ is the enclosed mass that forms a black hole.  The corresponding minimum angular momentum ($j_{\rm crit}$)is:  
\begin{equation}
    j_{\rm crit} = \sqrt{6} G M_{\rm BH}/c
\end{equation}
For a 5\,M$_\odot$ black hole, this critical specific angular momentum is $5.4\times10^{16} \, {\rm cm^2 s^{-1}}$.  As we shall see in figure~\ref{fig:rvmf}, the mass enclosed within $10^{10} {\rm \, cm}$ for BH-forming stars exceeds 5\,M$_\odot$.  For nearly all of our tidally-locked stars, the angular momentum at $10^{10} {\rm \, cm}$ is insufficient to produce a disk.  The specific angular momentum increases with radius squared, so the outer layers in some cases will produce a disk.  The stars with the highest angular momentum are a subset of the collapsing secondaries in tightly-bound orbits.  These stars are likely to form disks around low-mass BH cores and are the most-likely to form strong jets (see equation~\ref{eq:lgrb}).

To determine the radius (and mass) at which the infalling material will form a disk, we must first study the structure of the star at collapse.  Figure~\ref{fig:rvmf} shows the radius versus enclosed mass for different black-hole forming progenitors using 3 different stellar evolution codes:  KEPLER~\citep{2002RvMP...74.1015W}, GENEC~\citep{2000A&A...361..159M}, and MESA~\citep{2011ApJS..192....3P}.  The MESA and KEPLER calculations follow the evolution through silicon core burning and to stellar collapse.  The GENEC code does not model the star through silicon burning and, not surprisingly, is slightly more extended than the KEPLER and MESA simulations in the inner core (within 4\,M$_\odot$).  Beyond the inner core, the results vary between different stellar mass, rotation and code used.  Given the uncertainties in the models, we use an approximate fit for the enclosed mass above 2\,M$_\odot$ ($M_{\rm enc}$) to capture the features of this structure to calculate the angular momentum of the tidally-locked star:
\begin{equation}
    M_{\rm enc} = 7  (log_{\rm 10} (r+10^9) - 9) M_\odot
\end{equation}
where $r$ is the stellar radius in cm.

\begin{figure}
\includegraphics[scale=0.32,angle=0]{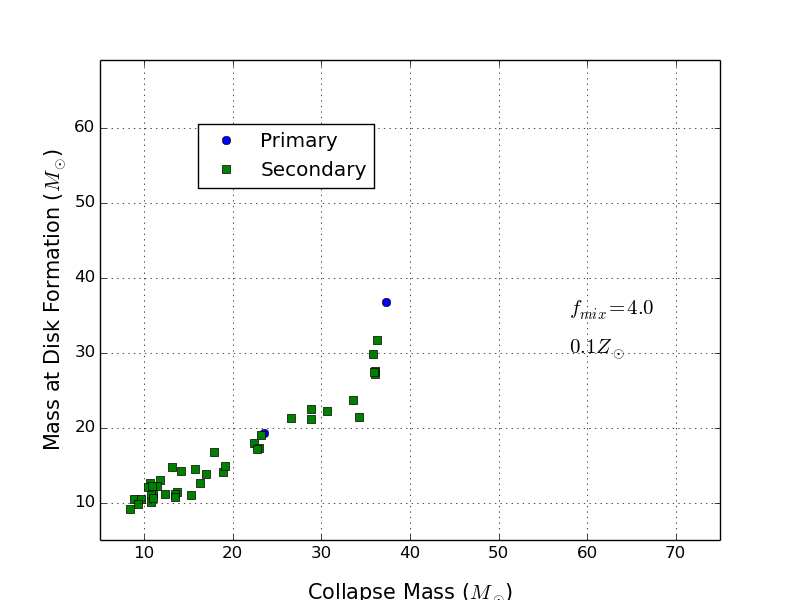}
\includegraphics[scale=0.32,angle=0]{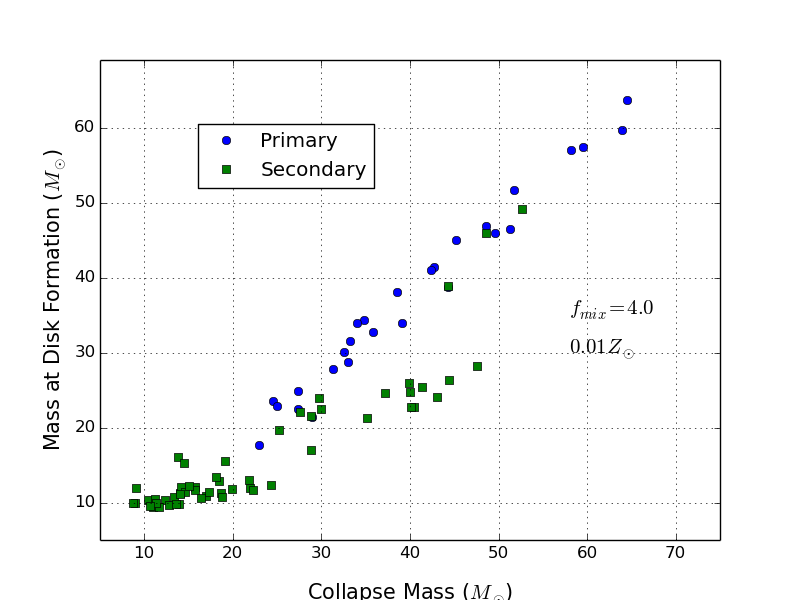}
\includegraphics[scale=0.32,angle=0]{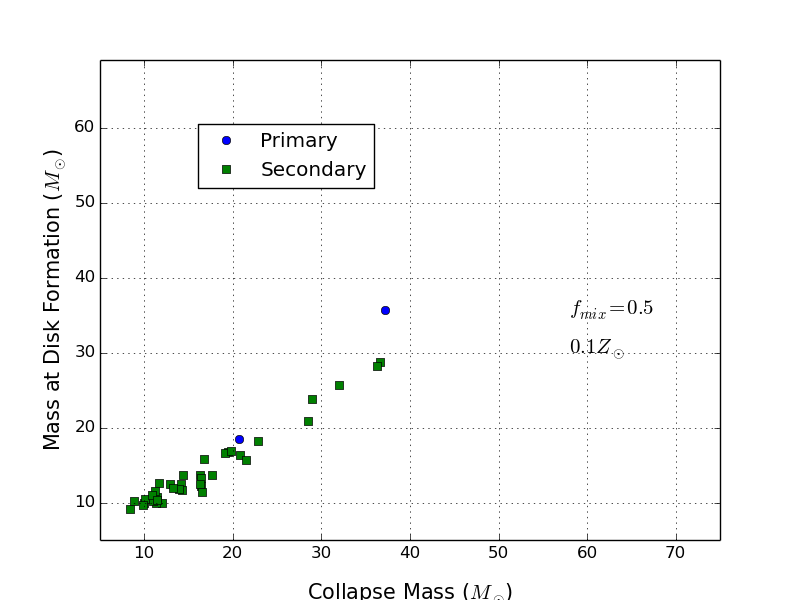}
\includegraphics[scale=0.32,angle=0]{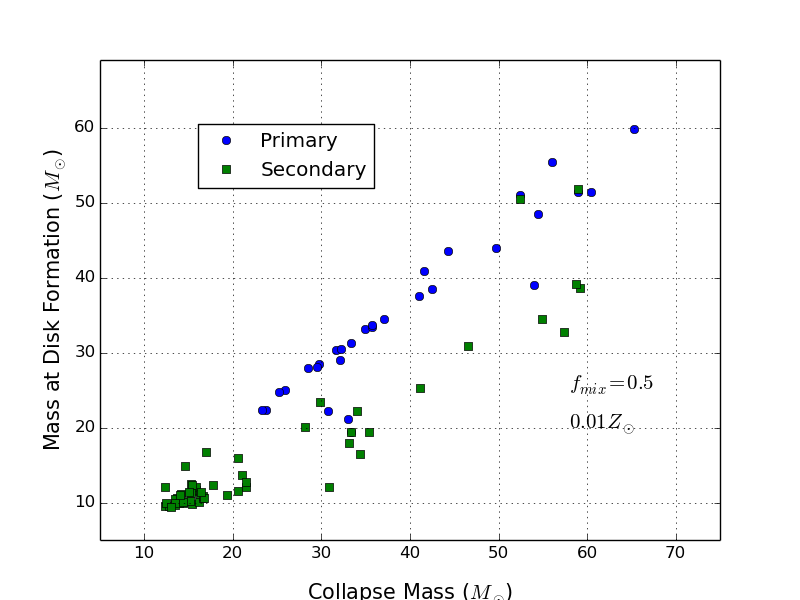}
\caption{Mass of remnant at the formation of the disk.  The final black hole mass will lie somewhere between this mass and the mass of the star at collapse.  The 4 panels show the {\tt StarTrack} results at two metallicities (0.01,0.1\,Z$_\odot$) and two mixing parameters ($f_{\rm mix}=0.5,4.0$).} 
\label{fig:mremrot}
\end{figure}

With this fit to the stellar model, we can use the spins and mass/radius relation of the stars to determine the position where the angular momentum exceeds that needed to form a disk for the enclosed mass.  Many of our systems are not spinning rapidly enough to form a disk.  Figure~\ref{fig:mremrot} shows the enclosed mass when the angular momentum from tidal locking causes the star to form a disk as a function of collapse mass for those systems that do form a disk (7-10\% of the binaries in our 4 population synthesis models).  The strength of the jet produced in these disks decreases rapidly with the mass of the interior compact remnant (see equation~\ref{eq:lgrb}).  Only 1\% of our binary systems produce black holes less massive than 10\,M$_\odot$ when the disk forms.  Because the power of the jet is proportional to the energy in the disk and the disk energy drops as the event horizon increases (with remnant mass), it is difficult to make a strong gamma-ray burst jet with more massive black holes.  As such, only 1\% of our models would produce strong collapsar Gamma-Ray Bursts. None of our systems had enough angular momentum to produce a disk around the core when the core was less than 3\,M$_\odot$ and these systems would not produce rotationally-powered supernovae.

\section{Conclusions}

In this paper, we build on our latest understanding of stellar evolution and the core-collapse supernova engine to study the compact remnant distribution.  We summarize the primary results here:
\begin{itemize}
    \item State-of-the-art simulations can not resolve the growth of convection.  Although common RANS models used to capture this subgrid convection agree on the short growth times and the dominate regions of convective instabilities, we demonstrate that the relative strengths depend on the leading terms in these models.
    \item We demonstrate that the convective growth time can produce a range of solutions for the remnant versus CO core mass.  The number of remnants within the NS/BH mass gap region depends on the growth timescales.  We provide a simple parameterized formula for these remnant masses.  
    \item We discussed 2 forms of stochasticity in the remnant mass distribution caused by:  the first captures the growth time variation that can be caused by asphericities in the collapsing core that then seed the convection. The latter captures the dramatic structural differences seen in some stars (e.g. small differences in shell burning lead to large changes in pre-collapse star structure).  Although stochasticity can alter the fate of an individual system, its effect on large populations is much less important.
    \item Although evidence continues to grow that the standard engine behind supernovae is from the convective model studied here, rapidly rotating models can explode via alternative mechanisms.  We studied the effect of rapid rotation on the black hole remnant mass distribution and its potential to fill in the pair-instability mass gap.
    \item We studied the effect of rotation on the black hole mass distribution if rapid rotation in stars is only produced through tidal locking using recent population synthesis models.  In such a spin-up scenario, roughly 10\% of stars in close binaries achieve sufficiently high angular momenta to form a disk.  Only 1\% of these stars form disks around remnants with masses less than 10\,M$_\odot$.  If this is the only way to produce fast-spinning stars, rotationally-driven explosions are common in only a small fraction of BH-forming stars and will not produce NSs.
\end{itemize}

The remnant mass distribution can be used to help us better constrain the growth timescales of convection in supernovae, reducing this key uncertainty in our understanding of the supernova engine.  Upcoming measurements both through lensing~\citep{2021arXiv210713697M} and gravitational wave observations of mergers with next-generation detectors~\citep{2021ApJ...913L...5N} are set to grow the number (and fidelity) of remnant distributions.  Coupled with better supernova calculations and binary studies (e.g. Olejak, in preparation), we can use these observations to drive breakthroughs in our understanding of the supernova engine and its implications on all of the astrophysics affected by these explosions.

\begin{acknowledgements}

We would like to thank Eric Burns, Mathieu Renzo, Lieke van Son and an anonymous referee for useful comments improving this paper.  The work by CLF was supported by the US Department of Energy through the Los Alamos National Laboratory. Los Alamos National Laboratory is operated by Triad National Security, LLC, for the National Nuclear Security Administration of U.S.\ Department of Energy (Contract No.\ 89233218CNA000001).f  KB and AO acknowledge support from the Polish National Science Center (NCN) grant Maestro (2018/30/A/ST9/00050)  

\end{acknowledgements}

\bibliography{refs}{}
\bibliographystyle{aasjournal}



\end{document}